\documentclass[aps,prl,twocolumn,superscriptaddress]{revtex4-1}

\usepackage[utf8]{inputenc}
\usepackage[english]{babel}
\usepackage{amsmath}
\usepackage{amsfonts}
\usepackage{amssymb}
\usepackage{makeidx}
\usepackage{graphicx}
\usepackage{lmodern}
\usepackage{kpfonts}
\usepackage[left=1.5cm,right=1.5cm,top=2cm,bottom=2cm]{geometry}

\usepackage{xcolor}
\usepackage{hyperref}
\hypersetup{linkbordercolor=cyan}

\usepackage{amsmath}

\usepackage{verbatim}
\usepackage{caption}
\usepackage{subcaption}
\usepackage{tabularx}
\usepackage{floatrow}

\usepackage{sidecap}
\usepackage{epstopdf}
\usepackage[margin=0.2cm]{subcaption}
\usepackage[toc,page]{appendix}
\usepackage{xcolor} 
\usepackage[export]{adjustbox}
\graphicspath{
{C:/Users/Utilisateur/Documents/Uni/these/latex/images/}
{C:/Users/Utilisateur/Dropbox/images/}
{C:/Users/Utilisateur/Dropbox/images/Langer/isotropic_annihilation/}
{C:/Users/Utilisateur/Dropbox/images/Langer/boundary_annihilation/}
{C:/Users/Utilisateur/Dropbox/images/Langer/2skyrmions_annihilation/}
{C:/Users/louise/Dropbox/images/}
{C:/Users/louise/Dropbox/images/screencaps/}
{C:/Users/louise/Dropbox/images/ffs/}
{C:/Users/louise/Dropbox/images/Langer/isotropic_annihilation/}
{C:/Users/louise/Dropbox/images/Langer/boundary_annihilation/}
{C:/Users/louise/Dropbox/images/Langer/2skyrmions_annihilation/}
{C:/Users/louise/Dropbox/images/Langer/}
{C:/Users/louise/Dropbox/images/Langer/non_mag_defect/}
{C:/Users/louise/Dropbox/images/Langer/blochsk_collapse/}
{C:/Users/louise/Dropbox/images/Langer/antisk_collapse/}
{C:/Users/louise/Dropbox/images/frustrated/sk_anni/}
{C:/Users/louise/Dropbox/images/frustrated/antisk_anni/}
{C:/Users/louise/Dropbox/images/frustrated/zoology/}
{C:/Users/louise/Dropbox/images/frustrated/Q2Q1/}
{C:/Users/louise/Dropbox/images/frustrated/Q2_Q1_Q1/}
{/home/desplat/Dropbox/images/Langer/boundary_annihilation/}
{/home/desplat/Dropbox/images/Langer/isotropic_annihilation/}
{/home/desplat/Dropbox/images/Langer/2skyrmions_annihilation/}
{/home/desplat/Dropbox/writing/latex/images/}
{/home/desplat/Dropbox/images/Langer/}
{/home/louise/Dropbox/writing/latex/images/}
{/home/louise/Dropbox/images/}
{/home/louise/Dropbox/images/screencaps/}
{/home/louise/Dropbox/images/Langer/isotropic_annihilation/}
{/home/louise/Dropbox/images/Langer/iso_collapse_SP2/}
{/home/louise/Dropbox/images/Langer/boundary_annihilation/}
{/home/louise/Dropbox/images/Langer/2skyrmions_annihilation/}
{/home/louise/Dropbox/images/Langer/antisk_collapse/}
{/home/louise/Dropbox/images/Langer/blochsk_collapse/}
{/home/louise/Dropbox/images/Langer/non_mag_defect/}
{/home/louise/Dropbox/images/Langer/}
{/home/louise/Dropbox/images/Langer/curved_boundary/}
{/home/louise/Dropbox/images/frustrated/sk_anni/}
{/home/louise/Dropbox/images/frustrated/antisk_anni/}
{/home/louise/Dropbox/images/frustrated/zoology/}
{/home/louise/Dropbox/images/frustrated/Q2Q1/}
{/home/louise/Dropbox/images/frustrated/Q2_Q1_Q1/}
{/home/louise/Dropbox/images/ffs/}
}

\usepackage{multirow}

\begin{document}

\title{Path sampling for lifetimes of metastable magnetic skyrmions\\ and direct comparison with Kramers' method}%

\author{L. Desplat}%
\email{louise.desplat@gmail.com}
\affiliation{Centre for Nanoscience and Nanotechnology, CNRS, Université Paris-Sud, Université Paris-Saclay, 91120 Palaiseau, France}
\affiliation{SUPA School of Physics and Astronomy, University of Glasgow, G12 8QQ Glasgow, United Kingdom}

 \author{ C. Vogler}
\affiliation{Christian Doppler Laboratory, Physics of Functional Materials, Faculty of Physics, University of Vienna, 1090 Vienna, Austria}

\author{J.-V. Kim}
\affiliation{Centre for Nanoscience and Nanotechnology, CNRS, Université Paris-Sud, Université Paris-Saclay, 91120 Palaiseau, France}

\author{ R. L. Stamps}
\affiliation{SUPA School of Physics and Astronomy, University of Glasgow, G12 8QQ Glasgow, United Kingdom}
\affiliation{Department of Physics and Astronomy, University of Manitoba, Winnipeg, Manitoba, R3T 2N2 Canada}

 \author{ D. Suess}
\affiliation{Christian Doppler Laboratory, Physics of Functional Materials, Faculty of Physics, University of Vienna, 1090 Vienna, Austria}

\date{\today}

\begin{abstract}
We perform a direct comparison between Kramers' method in many dimensions -- i.e., Langer's theory -- adapted to magnetic spin systems, and a path sampling method in the form of  forward flux sampling, as a means to compute collapse rates of metastable magnetic skyrmions. We show that a good agreement is obtained between the two methods. 
We report variations of the attempt frequency associated with skyrmion collapse by three to four orders of magnitude when varying the applied magnetic field by 5$\%$ of the exchange strength, which confirms the existence of a strong entropic contribution to the lifetime of skyrmions. This demonstrates that in complex systems, the knowledge of the rate prefactor, in addition to the internal energy barrier, is essential in order to properly estimate a lifetime.  
\end{abstract}

\maketitle


\paragraph*{}The rate of decay of metastable states is an ubiquitous
problem in physics. Thermal activation processes
across an energy barrier are found within fields as diverse
as solid state physics (Josephson junctions),
chemical reactions, electrical circuit theory (phase-locked
loops), laser physics, and magnetization switching in ferromagnets
\cite{coffey,risken1996fokker}. In the context of magnetic data storage,
information is stored in the form of 0 and 1 bits, corresponding
to uniformly magnetized grains pointing along
opposite directions. New challenges arise in the necessity
to design small magnetic structures capable of retaining
a given state against fluctuations for an average lifetime of 10 years at room temperature \cite{wild2017entropy}. The ability to precisely predict that lifetime is therefore crucial.  The rate of such thermally activated processes can be described by an Arrhenius law \cite{hanggi1990reaction},
\begin{equation}\label{eq:arrhenius}
k = \tau^{-1}= f_0 e^{-\beta \Delta E},
\end{equation}
in which $\beta=(k_BT)^{-1}$, $\Delta E$ is the internal energy barrier, and the prefactor $f_0$, commonly referred to as the attempt frequency, corresponds to a fundamental fluctuation rate. Estimating the stability of magnetic structures is often synonymous with accessing internal energy barriers, while assuming a typical value of the prefactor in the GHz range \cite{weller1999thermal,chen2010advances,lederman1994measurement,cortes2017thermal}.  Hence, it is generally accepted that $\beta \Delta E \sim 50$ at room temperature is a sufficient and necessary condition in order to achieve the desired stability.

In recent years, magnetic skyrmions \cite{bogdanov1989thermodynamically,bogdanov1994thermodynamically} have emerged as potential candidates for spintronics applications in data storage and logic devices \cite{fert2013skyrmions,iwasaki2013current,zhou2014reversible,muller2017magnetic,hsu2017electric}.
Magnetic skyrmions are particle-like spin textures carrying an integer topological charge. They are stabilized by the introduction of a characteristic lengthscale in a system via competing interactions. In particular, the existence of chiral skyrmions in non-centrosymmetric bulk magnets and thin magnetic films with broken inversion symmetry is made possible by the Dzyaloshinkii-Moriya interaction (DMI) \cite{dzyaloshinskii,moriya,fert1980role,crepieux1998dzyaloshinsky}. Isolated skyrmions typically live on the ferromagnetic (FM) background as metastable excitations, but, under the effect of thermal fluctuations, will eventually collapse back to the uniformly magnetized state. The problem of their thermal stability has so far yielded vastly different theoretical predictions depending on the system of interest \cite{rohart2016path,desplat2018thermal,bessarab2018annihilation}, particularly concerning the order of magnitude of the attempt frequency. Experimentally, extreme variations of $f_0$ were observed for small variations of the applied magnetic field in the case of the decay of the skyrmion lattice \cite{wild2017entropy}. The apparent lack of consensus between the results is in part due to the difficulty in calculating rate constants  of \textit{rare} events.  For structures with lifetimes well above the nanosecond range, direct Langevin simulations \cite{garcia1998langevin}, where one integrates the stochastic dynamics of the spin system at each timestep, becomes unrealistic.
In that case, a possible approach is the use of a form of reaction rate theory \cite{hanggi1990reaction,coffey}, which allows a direct calculation of the rate prefactor by considering details of the fluctuations about the metastable state $A$ and the saddle point (SP) $S$  along the reaction coordinate.  
Numerical implementations of this method \cite{suess2011calculation,fiedler2012direct} combined with a path finding scheme \cite{dittrich2002path,gneb} have previously been used to obtain lifetimes of magnetic skyrmions \cite{bessarab2018annihilation,desplat2018thermal,haldar2018first,von2019skyrmion}.
While this is a computationally optimal solution, reaction rate theory is based on many assumptions concerning the damping regime, the energy landscape, and the density of states of the system. Additionally, whenever we are faced with several mechanisms for a single process, we can only assume that the mechanisms are independent in order for the rates to add up, which may not hold. We are also faced with the questions of whether higher-order saddle points should contribute to the rate, and whether Eigenmodes with small Eigenvalues should be treated as Goldstone modes. An alternative method is therefore required in order to validate previous results. For that purpose, we turn to forward flux sampling (FFS) \cite{allen2004sampling,allen2006simulating,allen2006forward,allen2009forward,borrero2008optimizing}. FFS is a path sampling method that was initially developed to simulate rare switching events in biochemical networks. It has since then been applied to a wide range of rare event problems \cite{allen2009forward}. In particular, it was used to obtain magnetization switching rates in magnetic microstructures  \cite{vogler2013simulating,vogler2015calculating}. FFS was shown to be significantly more efficient than brute force direct Langevin simulations, while enabling an exploration of phase space free from assumptions. 
In this article, we demonstrate the application of the FFS method to the computation of collapse rates of metastable magnetic skyrmions far away from the system's boundaries, and we compare the results with predictions from Langer's theory, as well as with direct Langevin simulations whenever it can realistically be achieved.

\begin{figure}
\centering

		\includegraphics[width=.9\textwidth]{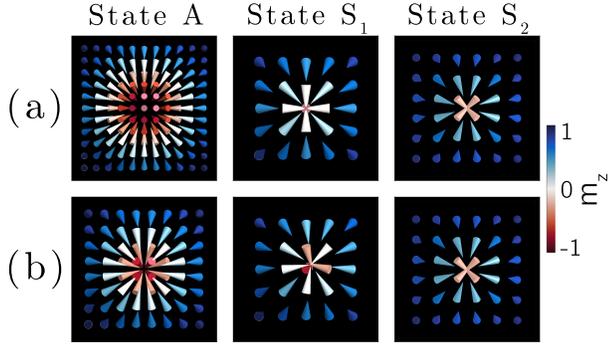}
	\caption{Spin maps (zoomed) of the metastable state $A$ and saddle points $S_1$ and $S_2$ of the skyrmion collapse for (a) $b_z =  0$, and (b) $b_z =  0.05$. $S_1$ corresponds to the skyrmion core centered on a lattice site, while $S_2$ corresponds to the core located at an interstitial point.}
\label{fig:collapse_paths}	
\end{figure}

We simulate $N$ magnetic spins of constant amplitude on a two-dimensional square lattice.  We use the classical Heisenberg model Hamiltonian,
\begin{equation}\label{eq:hamiltonian}	
	\begin{split}
& E = - J_{\text{ex}} \sum_{<ij>}  \mathbf{m}_i \cdot \mathbf{m}_j - \sum_{<ij>} \mathbf{D}_{ij} \cdot \big( \mathbf{m}_i \times \mathbf{m}_j \big)  \\
& -  K \sum_i m_{z,i}^2 -B_z M_s \sum_i m_{z,i}, \\
\end{split}
	\end{equation} 
where $J_{\text{ex}}$ is the strength of the isotropic exchange coupling, $\mathbf{D}_{ij}$ is the interfacial Dzyaloshinskii vector, $K$ is the perpendicular uniaxial anisotropy constant, $M_s$ is the saturation magnetization, and $B_z$ is the perpendicular applied magnetic field. Exchange interactions are restricted to first nearest neighbors.  We introduce the reduced parameters: $d = \lvert \mathbf{D}_{ij} \lvert / J_{\text{ex}} $; $k = K /J_{\text{ex}}$ ; $b_z = B_z M_s / J_{\text{ex}}$, and we set $(d,k)$ = (0.36, 0.4), which allows the existence of small Néel skyrmions solutions at zero-field that only span over about 7 lattice sites in diameter  (state $A$ in Fig. \ref{fig:collapse_paths}a) \cite{heo2016switching}. At low temperature, 
the skyrmions do not exhibit translational invariance on the lattice -- i.e., no Goldstone modes of zero-energy fluctuations -- but instead experience pinning at particular lattice positions. The applied field is oriented opposite to the skyrmion's core and has a destabilizing effect. The rest of the material parameters correspond to Pt/Co/AlO$_x$ samples \cite{miron2011fast,thiaville2012dynamics,rohart2013skyrmion} and are given in the Supplemental Material (SM) \cite{sm_ffs}. We simulate an infinite system by setting periodic boundary conditions, which eliminates cases where the skyrmion escapes at the edges \cite{uzdin2017effect,lobanov2016mechanism,bessarab2018annihilation,desplat2018thermal,cortes2017thermal}.


\begin{figure}
\begin{subfigure}[t]{.49\textwidth}	
		\includegraphics[width=1\textwidth]{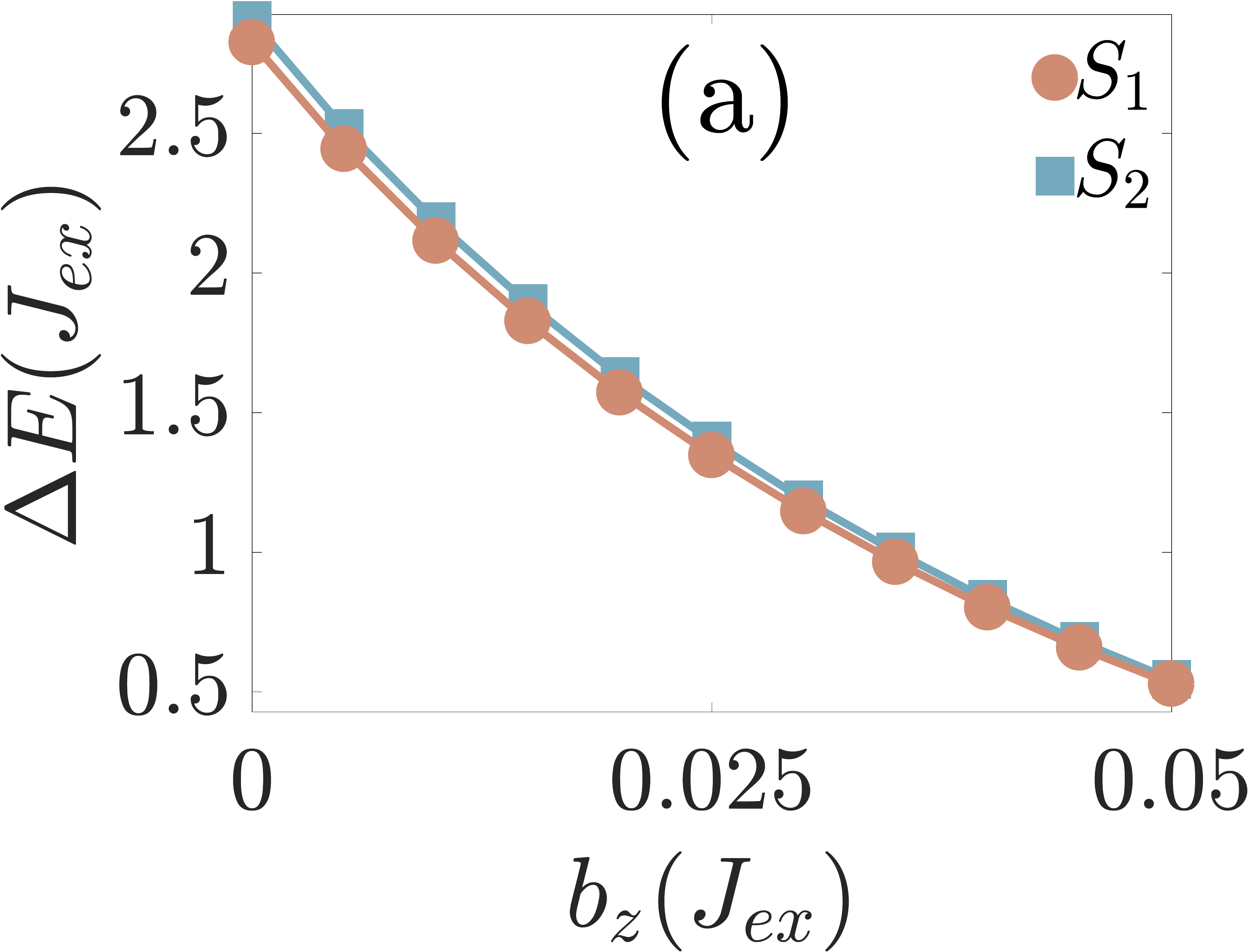}
	\label{fig:energy_barriers}
	\end{subfigure}	
	\begin{subfigure}[t]{.49\textwidth}		
\includegraphics[width=1\textwidth]{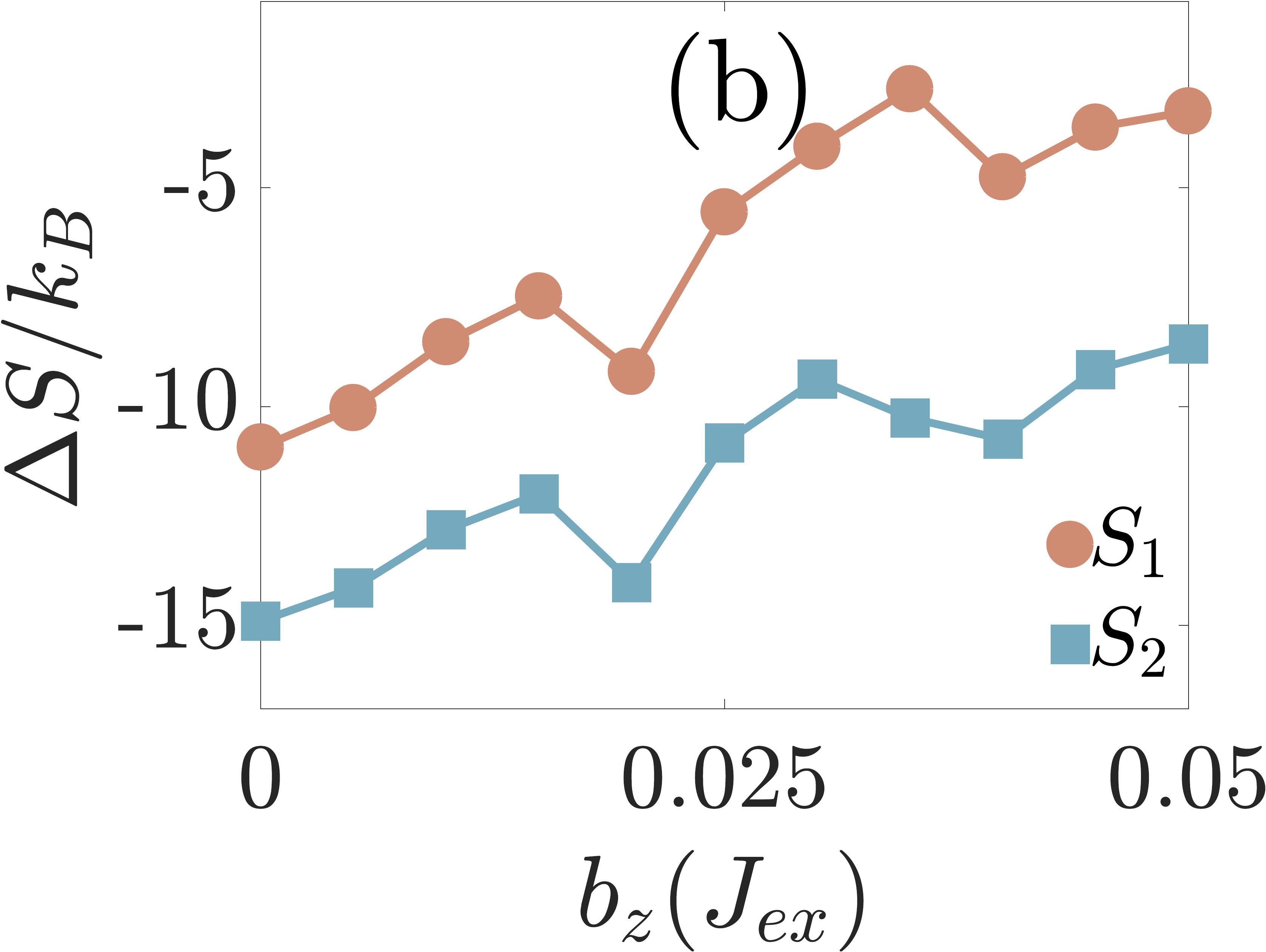}
		\label{fig:dS}
\end{subfigure}
\caption{Contributions to the change in free energy of the skyrmion upon reaching the saddle points $S_1$ and $S_2$, as a function of the applied field: (a) internal energy barrier, and (b) change in configurational entropy at $k_BT_{300}(=0.26 J_{\mathrm{ex}})$.\label{fig:deltaF}}
\end{figure}

\paragraph*{}We firstly relax the paths of minimum energy that lead to the skyrmion collapse on the energy landscape, and identify the saddle point along the path via the geodesic nudged elastic band method with a climbing image \cite{gneb,climbingimage}. The prefactor in Eq. (\ref{eq:arrhenius}), $f_0$, is then calculated via an extension of Kramers' method \cite{hanggi1990reaction} to many dimensions, namely, Langer's theory for the decay of metastable states \cite{langer} adapted to magnetic spin systems \cite{braun1994kramers,coffey,desplat2018thermal}. The theory is set in the intermediate-to-high damping regime. It yields no temperature dependence of $f_0$ if there are no Goldstone modes, or the same number of Goldstone modes at $A$ and $S$. Eq. (\ref{eq:arrhenius}) may be rewritten in terms of the change in Helmholtz free energy  $\Delta F$, 
\begin{equation}\label{eq:arrhenius_v2}
k = f_0' e^{-\beta \Delta F},
\end{equation}
where $f_0'$ is a new prefactor, and $\Delta F = \Delta E - T \Delta S$. $\Delta S$ corresponds to the change in configurational entropy undergone by the system upon reaching the saddle point. Details on the calculation of $f_0$ and $\Delta S$ can be found in \cite{sm_ffs,desplat2018thermal}. Note that $\Delta S$ is defined for stable modes of fluctuations, whereas $f_0$ takes into account both stable and unstable contributions.
We report two distinct collapse mechanisms. In one case, which we refer to as mechanism 1, the skyrmion shrinks in size while its core coincides with a lattice site, and the core-spin flips past the saddle point \cite{desplat2018thermal}. Alternatively, if mechanisms 2 is realized, the skyrmion core may shift to an interstitial position before uniformly shrinking \cite{gneb}. These two mechanisms involve distinct saddle points, that we respectively refer to as $S_1$ and $S_2$ [Fig. \ref{fig:collapse_paths}]. $S_2$ is found above $S_1$ on the energy surface, by advancing along the Eigenbasis coordinate associated with a translation mode. If the translational modes at the saddle points are not Goldstone modes, $S_1$ and $S_2$ should be treated as distinct states associated with different activation rates, namely $k_1$ and $k_2$.  When the metastable skyrmion is pinned at an interstitial position, the realization of mechanism 1 requires the core to firstly shift onto a lattice site [Fig. \ref{fig:collapse_paths}b]. The way the relaxed skyrmion sits on the lattice depends on its equilibrium size and its commensurability with the underlying lattice, such that there exists only one type of stable skyrmion state per field value, although the skyrmion can be indistinguishably located at either of the $N$ possible sites.  $S_1$ is a first-order saddle point, with a single unstable mode corresponding to the breathing of the skyrmion \cite{desplat2018thermal}. At $S_2$, three unstable modes are found -- the unstable breathing mode, and two unstable modes of translation -- resulting in a third-order SP. As we increase the field, the stable skyrmion size decreases \cite{bogdanov1994properties,wilson2014chiral,romming2015field}, and so do the internal energy barriers for both mechanisms, which we plot in Fig. \ref{fig:deltaF}a. In Fig.  \ref{fig:deltaF}b, we show the change in configurational entropy upon reaching the SP, which is found to become less negative as the field increases. Since $\Delta S < 0$ (entropic narrowing \cite{desplat2018thermal}), it is a stabilizing effect which lowers the attempt frequency. Lastly, we assume that the collapse processes are independent, so that the total rate of collapse is  $\tau^{-1}_{\text{Langer}}(T) = k_1(T) + k_2(T)$.


\begin{figure}
\centering
	\begin{subfigure}[t]{.49\textwidth}		
		\includegraphics[width=1\textwidth]{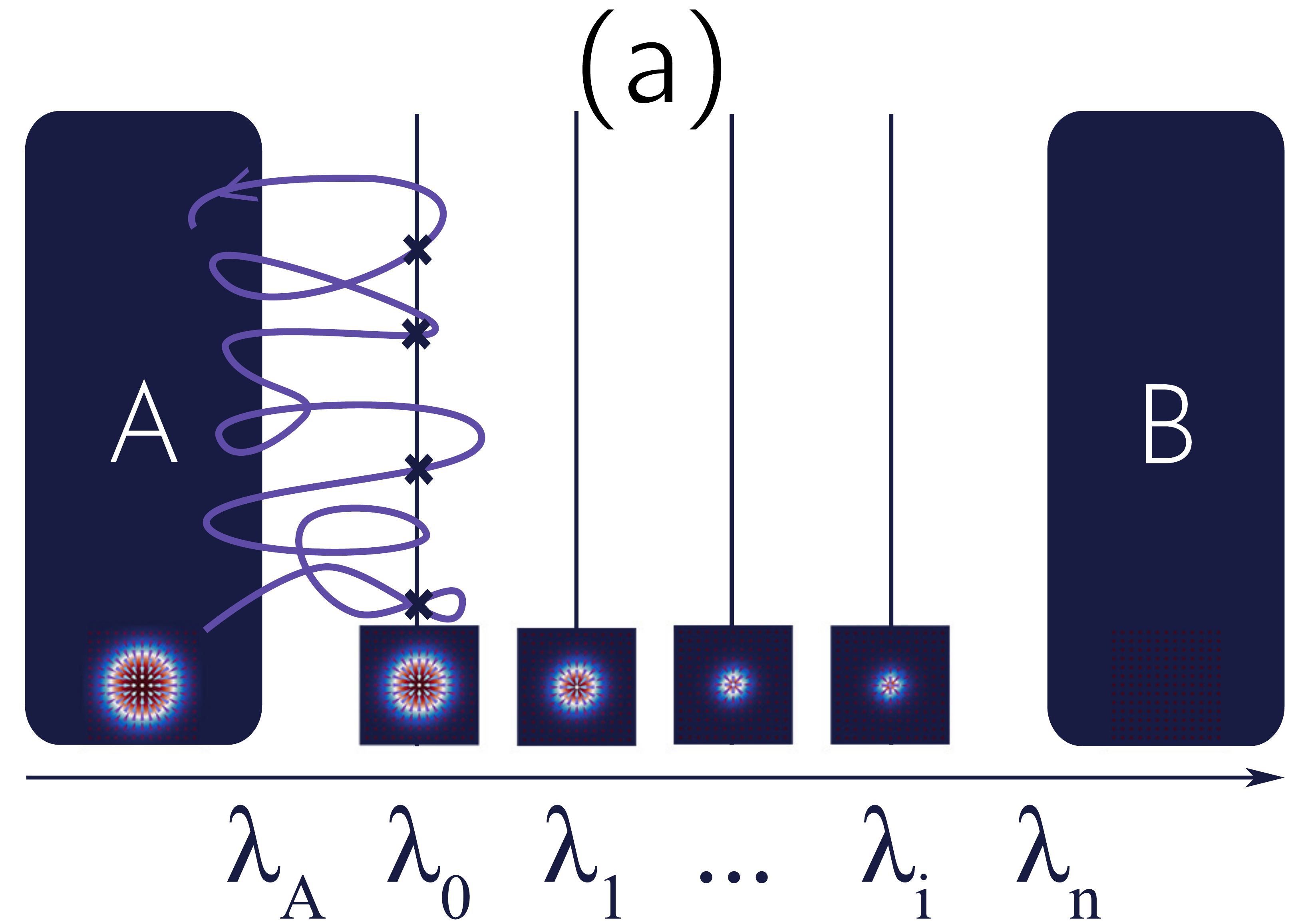}
		\label{fig:ffs_1}
	\end{subfigure}\hfill		
	\begin{subfigure}[t]{.49\textwidth}		
		\includegraphics[width=1\textwidth]{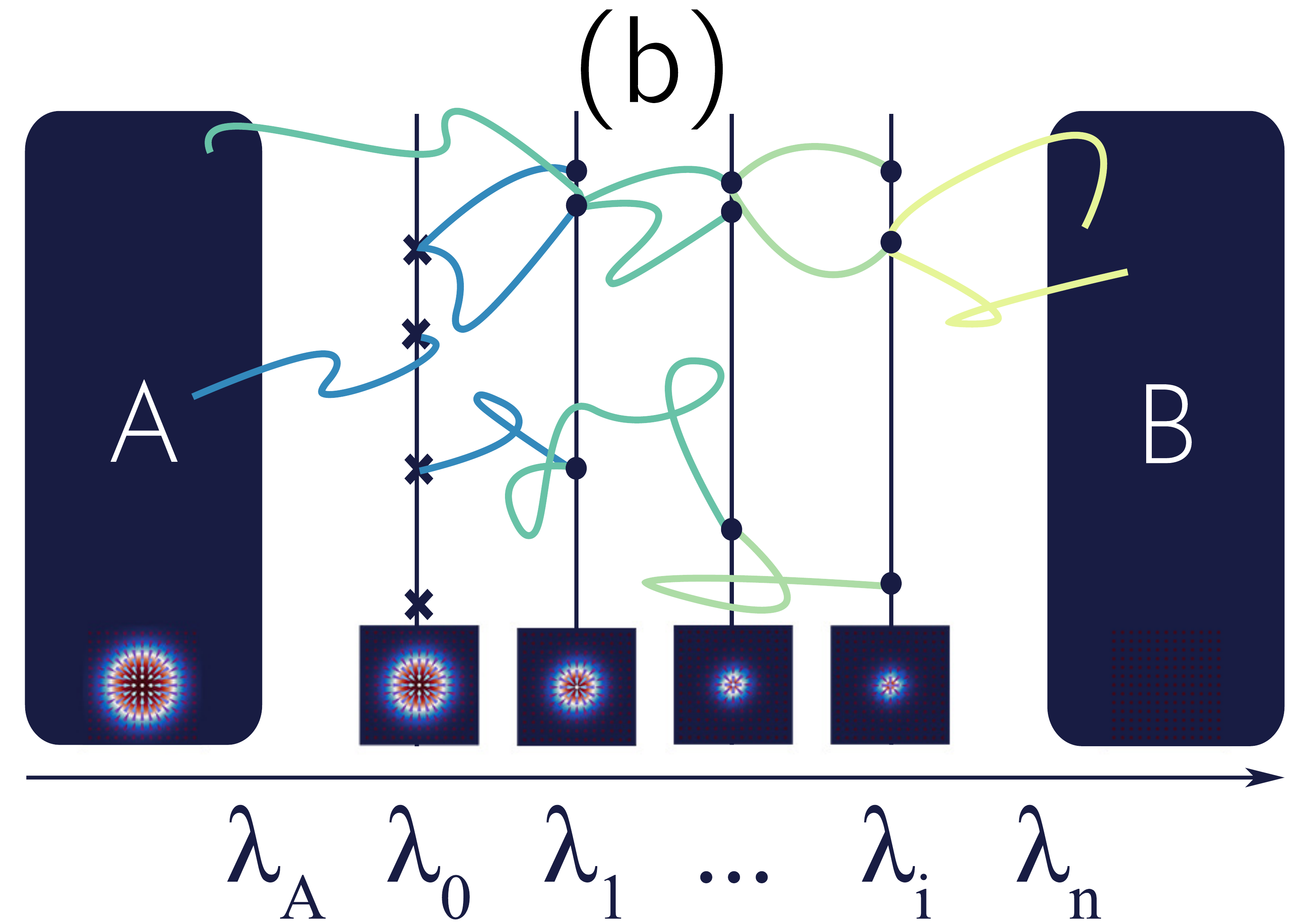}		\label{fig:ffs_2}
	\end{subfigure}
		\caption{Illustration of the FFS method, where the order parameter defining the interfaces $\{\lambda_i\}$ between  $A$ (metastable skyrmion) and $B$ (FM state) is chosen as the decreasing size of the skyrmion. (a) First simulation to compute the rate of crossing of $\lambda_0$. (b) Trial runs at subsequent interfaces. \label{fig:ffs}}
	
\end{figure}

\begin{figure}
\centering
			\begin{subfigure}[t]{.9\textwidth}
		\includegraphics[width=1\textwidth]{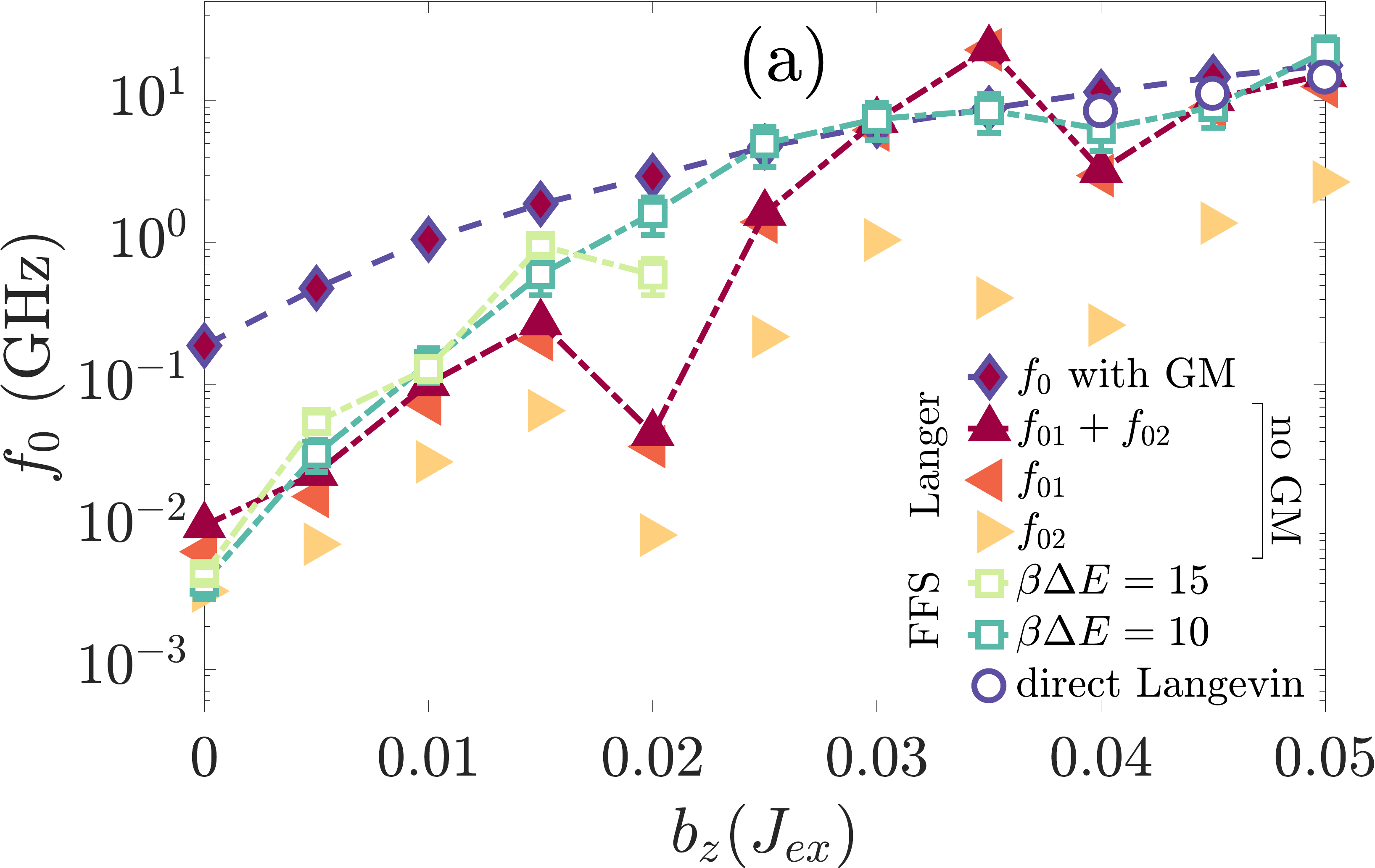} \label{fig:f0}
	\end{subfigure}\\[-1.3cm]
	
			\begin{subfigure}[t]{.9\textwidth}	
		\includegraphics[width=1\textwidth]{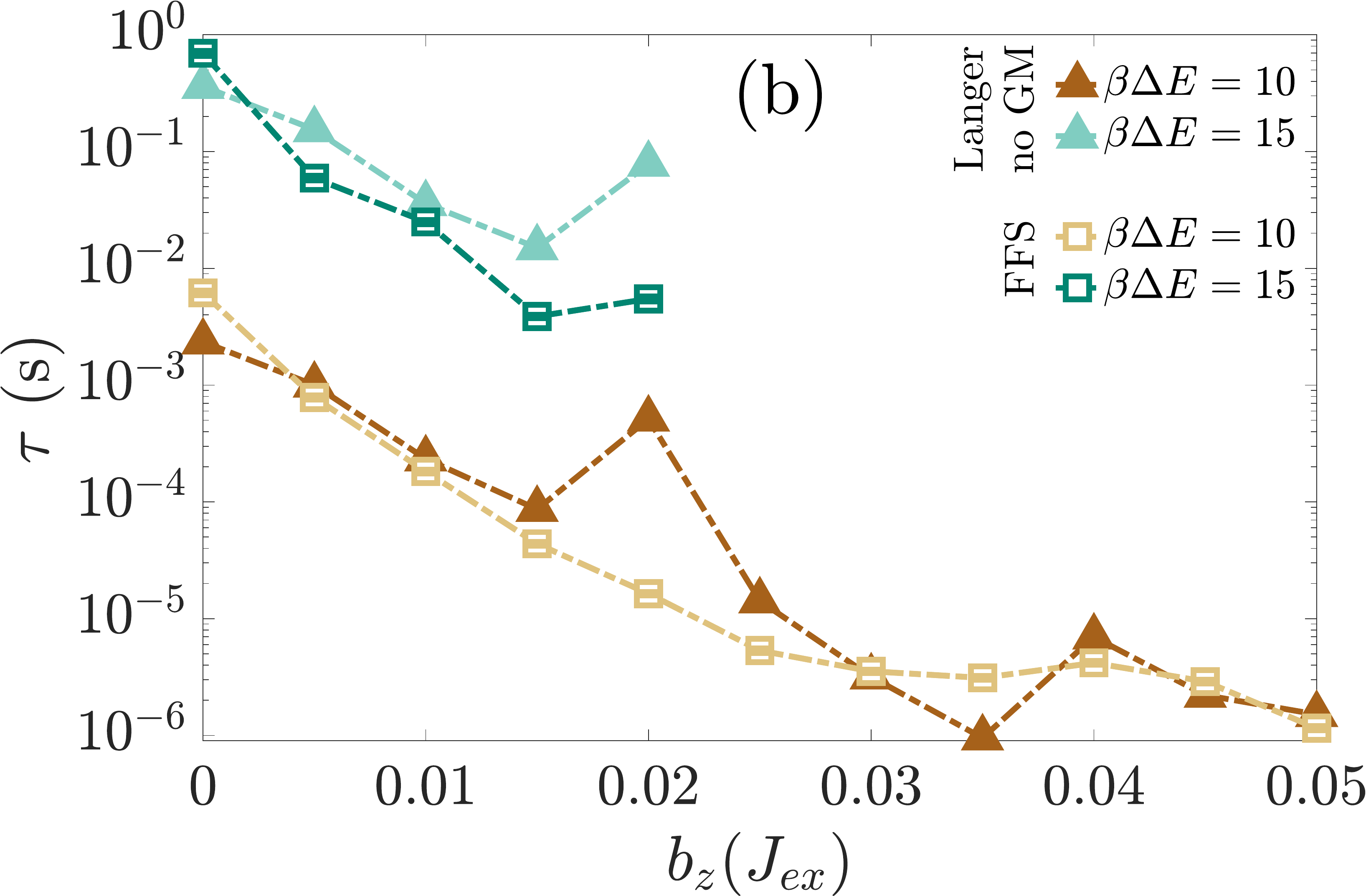}\label{fig:tau}
		\end{subfigure}
\caption{(a) Attempt frequency and (b) lifetime of the
skyrmion against collapse as a function of the reduced field calculated via Langer's theory with and without translational Goldstone modes (GM) and FFS. (a) also shows the result of direct Langevin simulations.\label{fig:ffs_plots}}
\end{figure}

\begin{figure}
			\includegraphics[width=.9\textwidth]{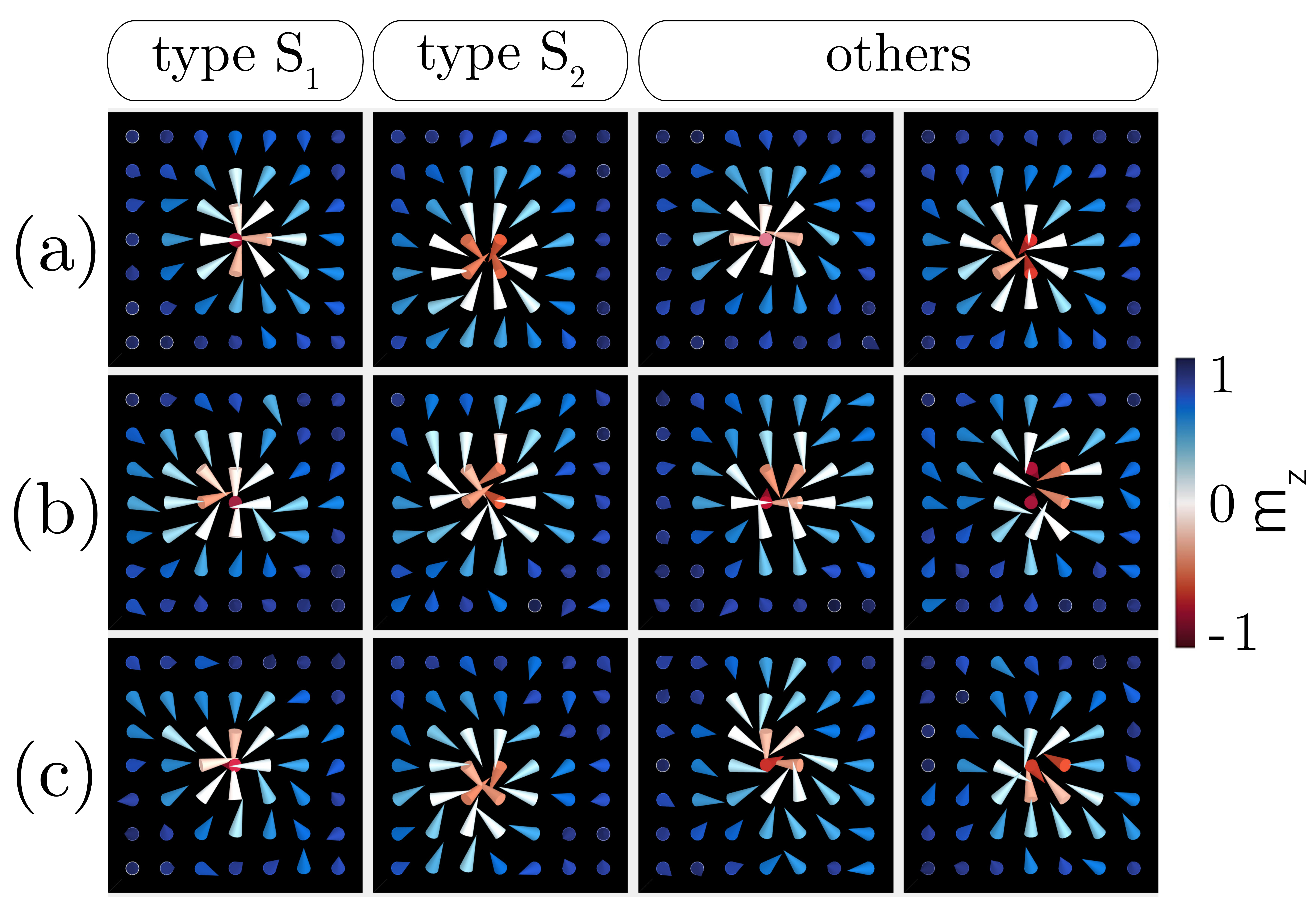}				
		\caption{Examples of (zoomed) spin configurations from FFS stored at the interface(s) at which the order-parameter equates that of a SP (a) at interface $\lambda(x=4)$, for $b_z = 0.05$ and $\beta \Delta E_1 = 10$ (b, c) for $b_z = 0$ and $\beta \Delta E_1 = 15$, at interfaces (b)  $\lambda(x=5)$, and (c)   $ \lambda(x=4)$.		\label{fig:ffs_configs}}
\end{figure}

\paragraph*{}Our aim is to compare the results of Langer's theory with collapse rates obtained from forward flux sampling. The FFS method generates trajectories between two (meta)stable states $A$ and $B$ in a ratchet-like manner without imposing any bias on the microscopic dynamics, which makes it well adapted for the simulation of rare events. Compared to other path sampling methods, it does not require prior knowledge of the density of states, which makes it suitable for nonequilibrium systems that do not obey detailed balance. It employs a set of $n (+1)$ nonintersecting interfaces in phase space to sample the transition path ensemble and compute a transition rate. The interfaces  $\{\lambda_A, \lambda_0 \dots \lambda_n = \lambda_B\}$ [Fig. \ref{fig:ffs}] are defined as iso-surfaces of a monotonically varying order parameter, $ \lambda (x_i) = \lambda_i$, such that $x_{i+1} > x_i$ or $x_{i+1} < x_i$ for all $i$. Any trajectory going from $A$ to $B$ must cross all the interfaces at least once.  The rate constant from $A$ to $B$ may be expressed as
\begin{equation}\label{eq:ffs_rate}
k_{\text{FFS}} = \Phi_{A,0} \prod_{i=0}^{n-1} P(\lambda_{i+1} \vert \lambda_i),
\end{equation}
in which $\Phi_{A,0}$ is the rate at which trajectories starting from region $A$ cross the first interface $\lambda_0$, and the conditional probabilities $P(\lambda_{i+1} \vert \lambda_i)$  correspond to the probability that a trajectory coming from $A$ that crossed $\lambda_i$ for the first time will cross $\lambda_{i+1}$ before returning to $A$. The protocol is illustrated in Fig. \ref{fig:ffs} and is as follows. First, a single Langevin simulation is started in state $A$ [Fig. \ref{fig:ffs}a]. Each time the system successfully exits region $A$ and crosses $\lambda_0$, its configuration is stored. The simulation ends after $N_0$ crossing events have been recorded, and the flux of trajectories out of $A$ is obtained by $ \Phi_{A,0}  = N_0 / \Delta t$, in which $\Delta t$ is the total simulated time. After that, a configuration stored at $\lambda_0$ is selected at random and used as a starting point for a new simulation [Fig. \ref{fig:ffs}b]. That new simulation ends when the trajectory either crosses $\lambda_1$, in which case the crossing configuration is stored, or the system returns to $A$. This procedure is repeated $M_0$ times. If $N_0^s$ trajectories successfully crossed $\lambda_1$,  we obtain $P(\lambda_{1} \mid \lambda_0) =  N_0^s / M_0$. One then proceeds analogously at subsequent interfaces. During the trial runs, Langevin simulations are carried out by integrating the system of stochastic Landau-Lifshitz-Gilbert equations at each timestep, by means of the stochastic Heun scheme \cite{garcia1998langevin}, for which details can be found in the SM \cite{sm_ffs}. To obtain the rate of collapse of a skyrmion, a natural choice of order parameter is the skyrmion size, where state $A$ is the equilibrium skyrmion size, and state $B$ -- corresponding to the FM state -- is associated to a zero-size  [Fig. \ref{fig:ffs}].
Arbitrarily, we consider that magnetic sites $\mathbf{m}_i$ ($i=1 \hdots N$) that satisfy $m_{z,i}\leq 0$  are part of the skyrmion, and we define the order-parameter $x$ as the (integer) number of sites inside the skyrmion. For values of the reduced field in $[0, 0.05]$, we compute a total collapse rate, $k_{\text{FFS}}$. We give the results  from FFS and Langer’s theory for the attempt frequency and the skyrmion lifetime against collapse in Figs. \ref{fig:ffs_plots}a and \ref{fig:ffs_plots}b.  Through each FFS run, we set the temperature such that $\beta \Delta E_1 =$  10, so that Langer's theory may hold \cite{coffey}. FFS runs are also carried out at $\beta \Delta E_1$ =  15 in the lower field region and yield very similar results, which shows that the attempt frequency has no significant $T$-dependence here.
Since the translation of the skyrmion costs little energy compared to $k_BT$, we also show Langer's result with a treatment of the translational modes as Goldstone modes \cite{buttiker1981nucleation,braun1994kramers,bessarab2018annihilation}. Details on the method can be found in the SM \cite{sm_ffs}. We find that this treatment results in $f_0$ being overestimated in the low field region, and, in this system, the best overall agreement between Langer and FFS is obtained without considering Goldstone modes. 
In Fig. \ref{fig:ffs_configs}, we show some examples of stored configurations at interfaces at which the order-parameter equates that of a saddle point. We report both $S_1$- and $S_2$-types of configurations, as well as some other configurations that don't clearly pertain to either category. This occurs because the crossing configurations correspond to an order parameter which is either equal to, or smaller than that of the SP, which does not imply that the configuration is in fact a SP.  
Under the effect of thermal fluctuations, the system does not usually cross the barrier exactly at the SP, but deviates from it by a more or less small amount. 
Lastly, at higher field values where $f_0$ is found within the GHz range, we also compute a collapse rate via direct Langevin simulations at $\beta_{300} \Delta E_1 \approx$ 2 - 3. Following a similar procedure to Ref. \cite{desplat2019paths}, we compute an average lifetime out of 400 collapses. The results are shown in Fig. \ref{fig:ffs_plots}a for $b_z \geq 0.04$ and match the FFS results.

We have thus validated the use of Langer's theory to obtain skyrmion collapse rates by means of forward flux sampling simulations. In the end, we find that due to a higher activation entropy, the path involving the third-order saddle point $S_2$ does not contribute significantly to the total rate. Nonetheless, the internal energy barriers associated with $S_2$ are almost the same as the ones associated with $S_1$, and configurations similar to $S_2$ are observed in FFS, so we could not justify neglecting it \textit{a priori}. We should also note that since the difference in internal energy between $S_1$ and $S_2$ is quite small (around 0.08 $J_{\mathrm{ex}}$ at zero field), the mechanisms are probably not completely decoupled. Since the skyrmions are coupled to the lattice, we observe lattice effects in the Langer approach, that manifest in non-monotonic variations of the entropic contribution [Fig. \ref{fig:deltaF}b], which are in turn found in the attempt frequency and the average lifetime [Fig. \ref{fig:ffs_plots}]. In FFS simulations, lattice effects are likely smoothed out by thermal fluctuations. Nevertheless, FFS shows that translational modes should not be treated as Goldstone modes in this case. FFS is a valuable tool, as it requires no prior assumptions on the system. With appropriate interface design, it could be used to treat problems that have not yet been successfully solved by reaction rate theory, such as the problem of skyrmion nucleation rates.
Most notably, FFS and Langer's theory both yield variations of the collapse rate prefactor by three to four orders of magnitude when the applied magnetic field varies by 0.05 $J_{\mathrm{ex}}$. This effect is due to the important entropic contribution, and implies that reaching the 10-year retention rate necessary for technological applications may require adequately tuning the attempt frequency, in addition to the energy barrier. This result is valid for magnetic skyrmions, but also applies to any (meta)stable state undergoing a consequent change in entropy upon reaching the transition state. Here, a decrease in entropy at the saddle point stabilizes the skyrmion state. This is directly linked to the skyrmion's internal modes \cite{desplat2018thermal}. Since the skyrmion size decreases with the applied field, we find that the bigger the skyrmion, the higher $\Delta E$, the more negative $\Delta S$, the stronger the stabilizing effect (see also Ref. \cite{von2019skyrmion}). On the other hand, for processes with high activation energies, the Meyer-Neldel compensation rule yields
a destabilizing, often large entropic contribution 
 \cite{meyer1937relation,yelon1990microscopic,yelon1992origin} (e.g. biological death rates, transport in semiconductors, decay of the skyrmion lattice, etc  \cite{rosenberg1971quantitative,peacock1982compensation,wild2017entropy,kamiya2010present,cooper2001heat}). These considerations underline the fact that, when estimating transition rates, one should not \textit{a priori} assume a characteristic value of $f_0$, and special care needs to be taken in its evaluation.

\begin{acknowledgments}
This work was supported by the Horizon 2020 Framework
Programme of the European Commission, under Grant agreement
No. 665095 (MAGicSky), and the Agence Nationale de la Recherche under Contract No. ANR-17-CE24-0025 (TOPSKY). The support from CD-Laboratory AMSEN (financed by the Austrian Federal Ministry of Economy, Family and Youth, the National Foundation for Research, Technology and Development) is acknowledged.  FFS simulations were performed on the Vienna Scientific Cluster (VSC).
\end{acknowledgments}
\bibliography{/home/louise/Dropbox/writing/latex/skyrmionbib}
 
\end{document}